\documentclass[10pt,letterpaper]{article}
\usepackage[top=0.85in,left=2.75in,footskip=0.75in]{geometry}

\usepackage[all]{xy}
\usepackage[dvips]{graphicx}
\usepackage{color}

\newcommand{\A}{\mathbf{A}}

\newcommand{\J}{\mathbf{J}}

\newcommand{\bsigma}{\boldsymbol{\sigma}}

\usepackage{amssymb}
\usepackage{amsfonts}
\usepackage{amsmath}
\usepackage[linesnumbered,ruled]{algorithm2e}
\usepackage{color}
\usepackage{caption}
\usepackage{subcaption}

\usepackage{changepage}

\usepackage[utf8x]{inputenc}

\usepackage{textcomp,marvosym}

\usepackage{fixltx2e}

\usepackage{amsmath,amssymb}

\usepackage{cite}

\usepackage{nameref,hyperref}

\usepackage[right]{lineno}

\usepackage{microtype}
\DisableLigatures[f]{encoding = *, family = * }

\raggedright
\usepackage[aboveskip=1pt,labelfont=bf,labelsep=period,justification=raggedright,singlelinecheck=off]{caption}
\renewcommand{\figurename}{Fig}

\makeatletter
\renewcommand{\@biblabel}[1]{\quad#1.}
\makeatother

\date{}

\usepackage{lastpage,fancyhdr,graphicx}
\usepackage{epstopdf}
\fancyheadoffset[L]{2.25in}
\fancyfootoffset[L]{2.25in}
\begin{document}
\vspace*{0.2in}

\begin{flushleft}
{\Large
\textbf\newline{The Energy Landscape of Neurophysiological Activity Implicit in Brain Network Structure} 
}
\newline
\\

Shi Gu\textsuperscript{1,2},
Matthew Cieslak\textsuperscript{3}, 
Benjamin Baird\textsuperscript{4},
Sarah F. Muldoon\textsuperscript{5},
Scott T. Grafton\textsuperscript{3},
Fabio Pasqualetti\textsuperscript{6},
 Danielle S. Bassett\textsuperscript{2,7,*}
\\

\bigskip
\textbf{1} Applied Mathematics and Computational Science, University of Pennsylvania, Philadelphia, PA, 19104 USA
\\
\textbf{2} Department of Bioengineering, University of Pennsylvania, Philadelphia, PA, 19104 USA
\\
\textbf{3} Department of Psychological and Brain Sciences, University of California, Santa Barbara, CA, 93106 USA
\\
\textbf{4} Center for Sleep and Consciousness, University of Wisconsin - Madison, Madison, WI 53706
\\
\textbf{5} Department of Mathematics and CDSE Program, University at Buffalo, SUNY, Buffalo, NY 14260
\\
\textbf{6} Department of Mechanical Engineering, University of California, Riverside, CA, 92521 USA
\\
\textbf{7} Department of Electrical \& Systems Engineering, University of Pennsylvania, Philadelphia, PA, 19104 USA
\\
\textbf{*} To whom correspondence should be addressed: dsb@seas.upenn.edu.
\\
\bigskip

%
%






\end{flushleft}
\section*{Abstract}
A critical mystery in neuroscience lies in determining how anatomical structure impacts the complex functional dynamics of human 
thought. How does large-scale brain circuitry constrain states of neuronal activity and transitions between those states? We address 
these questions using a maximum entropy model of brain dynamics informed by white matter tractography. We demonstrate that the 
most probable brain states -- characterized by minimal energy -- display common activation profiles across brain areas:  local 
spatially-contiguous sets of brain regions reminiscent of cognitive systems are co-activated frequently. The predicted activation rate 
of these systems is highly correlated with the observed activation rate measured in a separate resting state fMRI data set, validating 
the utility of the maximum entropy model in describing neurophysiologial dynamics. This approach also offers a formal notion of the 
energy of activity within a system, and the energy of activity shared between systems. We observe that within- and between-system 
energies cleanly separate cognitive systems into distinct categories, optimized for differential contributions to integrated 
\emph{versus} segregated function. These results support the notion that energetic and structural constraints circumscribe brain dynamics, offering novel insights into the roles that cognitive systems play in 
driving whole-brain activation patterns.

\section*{Author Summary}
How does the complex interconnection structure between large-scale brain regions impact how we think? Does this structure inform how we move between different thoughts or actions? We address these questions using a simple mathematical model of brain dynamics that is informed by empirical estimates of anatomical interconnection structure between brain regions. Our results suggest that while interconnection structure plays an important role in constraining and predicting brain dynamics, an additional layer of explanatory power is offered by considering energetic constraints on those dynamics. These data offer a richer landscape of mechanisms that enhance our understanding of how we may move from one thought to the next.


\section*{Introduction}

A human's adaptability to rapidly changing environments depends critically on the brain's ability to carefully control the time within 
(and transitions among) different states. Here, we use the term \emph{state} to refer to a pattern of activity across neurons or brain 
regions \cite{tang2012neural}. The recent era of brain mapping has beautifully demonstrated that the pattern of activity across the 
brain or portions thereof \cite{Mahmoudi2012} differs in different cognitive states \cite{Gazzaniga2013}. These variable patterns of 
activity have enabled the study of cognitive function \emph{via} the manipulation of distinct task elements \cite{Gazzaniga2013}, the 
combination of task elements \cite{Szameitat2011,Alavash2015}, or the temporal interleaving of task elements 
\cite{Ruge2013,Muhle2014}. Such methods for studying cognitive function are built on the traditional view of mental chronectomy 
\cite{Donders1969}, which suggests that brain states are additive and therefore separable in both space and time (although see 
\cite{Mattar2015} for a discussion of potential caveats).

Philosophically, the supposed separability and additivity of brain states suggests the presence of strong constraints on the patterns of 
activations that can be elicited by the human's environment. The two most common types of constraints studied in the literature are 
energetic constraints and structural constraints \cite{bullmore2012economy}. Energetic constraints refer to fundamental limits on the 
evolution \cite{Niven2008} or usage of neural systems \cite{Attwell2001}, which inform the costs of establishing and maintaining 
functional connections between anatomically distributed neurons \cite{Bassett2010}. Such constraints can be collectively studied 
within the broad theory of brain function posited by the free energy principal -- a notion drawn from statistical physics and 
information theory -- which states that the brain changes its state to minimize the free energy in neural activity 
\cite{friston2006free,friston2010free}. The posited preference for low energy states motivates an examination of the time within and 
transitions among local minimums of a predicted energy landscape of brain activity \cite{moreno2007noise,tsodyks1998neural}.

While energetic costs likely form critical constraints on functional brain dynamics, an arguably equally important influence is the underlying structure and anatomy linking brain areas. Intuitively, quickly changing the activity of two brain regions that are \emph{not} directly connected to one another by a structural pathway may be more challenging than changing the activity of two regions that \emph{are} directly connected \cite{achard2007efficiency,Bassett2010}.  Indeed, the role of structural connectivity in constraining and shaping brain dynamics has been the topic of intense neuroscientific inquiry in recent years \cite{honey2009predicting, honey2010can,Deco2012,Goni2014,becker2015}.  Evidence suggests that the pattern of connections between brain regions directly informs not only the ease with which the brain may control state transitions \cite{Gu2015}, but also the ease with which one can externally elicit a state transition using non-invasive neurostimulation \cite{Muldoon2016}.

While energy and anatomy both form critical constraints on brain dynamics, they have largely been studied in isolation, hampering an 
understanding of their collective influence. Here, we propose a novel framework that combines energetic and structural constraints 
on brain state dynamics in a free energy model explicitly informed by structural connectivity. Using this framework, we map out the 
predicted energy landscape of brain states, identify local minima in the energy landscape, and study the profile of activation patterns 
present in these minima.  Our approach offers fundamental insights into the distinct role that brain regions and larger cognitive 
systems play in distributing energy to enable cognitive function. Further, the results lay important groundwork for the 
study of energy landscapes in psychiatric disease and neurological disorders, where brain state transitions are known to be critically 
altered but mechanisms driving these alterations remain far from understood \cite{Ravizza2010,Wylie2010}.

\begin{figure}[!h]
\includegraphics[width=4in]{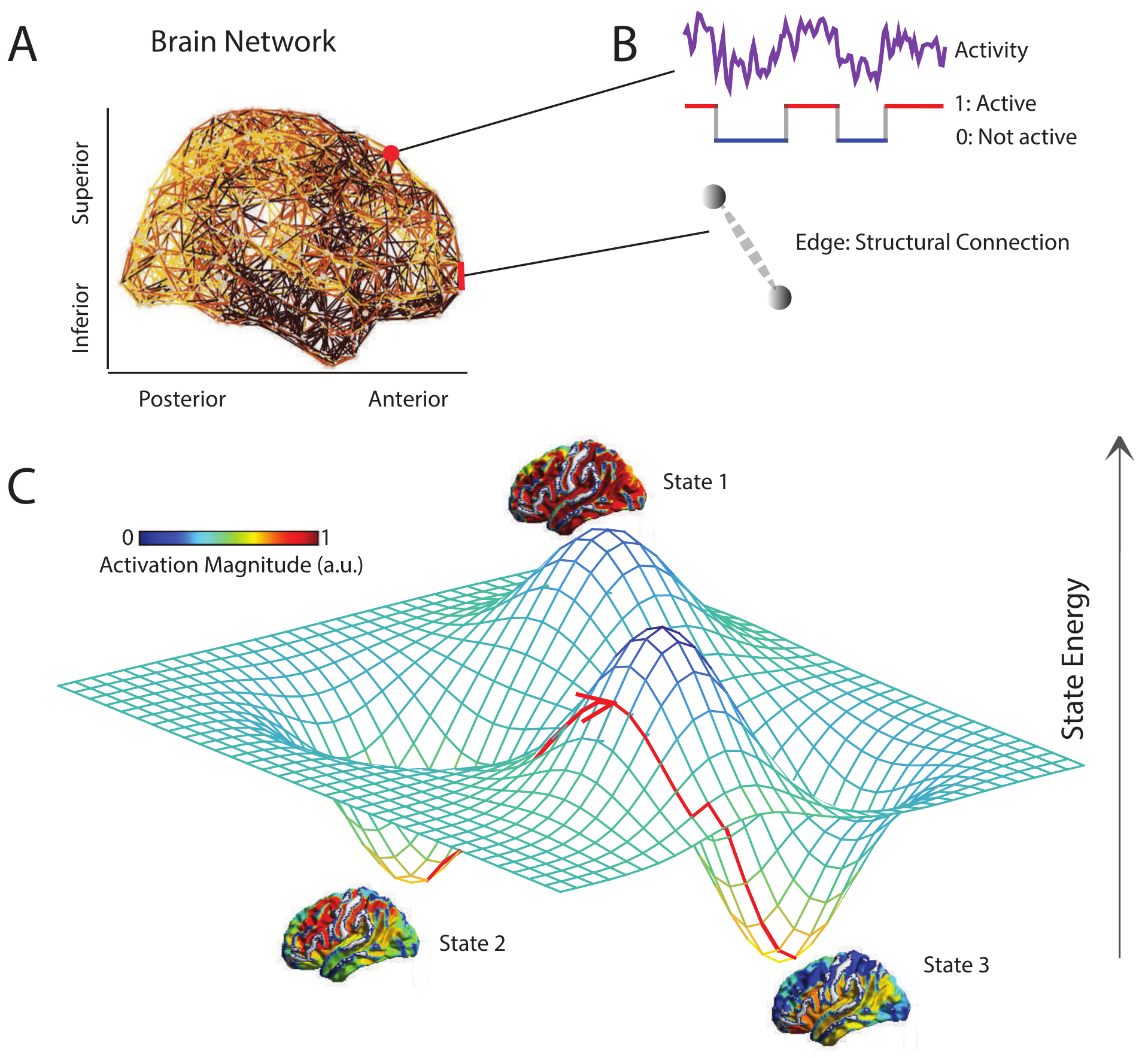}	
\caption{\textbf{Conceptual Schematic.} \emph{(A)} A weighted structural brain network represents the number of white matter 
streamlines connecting brain regions. \emph{(B)} We model each brain region as a binary object, being either active or inactive. 
\emph{(C)} Using a maximum entropy model, we infer the full landscape of predicted activity patterns -- binary vectors indicating the 
regions that are active and the regions that are not active -- as well as the energy of each pattern (or \emph{state}). We use this 
mathematical framework to identify and study local minima in the energy landscape: states predicted to form the foundational 
repertoire of brain function. }
 \label{fig1}
 \end{figure}

\section*{Materials and Methods}

\subsection*{Human DSI Data Acquisition and Preprocessing}
Diffusion spectrum images (DSI) were acquired for a total of $48$ subjects (mean age 22.6$\pm$5.1 years, 24 female, 2 left handed) along with a $T1$ weighted anatomical scan at each scanning session \cite{Cieslak2014}. Of these subjects, $41$ were scanned ones, $1$ was scanned twice, and $6$ were scanned three times, for a total of 61 scans. 

DSI scans sampled $257$ directions using a $Q5$ half shell acquisition scheme with a maximum $b$ value of $5000$ and an isotropic voxel size of $2.4$mm. We utilized an axial acquisition with the following parameters: $TR=11.4$s, $TE=138$ms, $51$ slices, FoV ($231$,$231$,$123$ mm). All participants volunteered with informed written consent in accordance with the Institutional Review Board/Human Subjects Committee, University of California, Santa Barbara.

DSI data were reconstructed in DSI Studio (www.dsi-studio.labsolver.org) using $q$-space diffeomorphic reconstruction (QSDR) \cite{Yeh2011}. QSDR first reconstructs diffusion weighted images in native space and computes the quantitative anisotropy (QA) in each voxel. These QA values are used to warp the brain to a template QA volume in MNI space using the SPM nonlinear registration algorithm. Once in MNI space, spin density functions were again reconstructed with a mean diffusion distance of $1.25$ mm using three fiber orientations per voxel. Fiber tracking was performed in DSI Studio with an angular cutoff of $55^{\circ}$, step size of $1.0$ mm, minimum length of $10$ mm, spin density function  smoothing of $0.0$, maximum length of $400$ mm and a QA threshold determined by DWI signal in the CSF. Deterministic fiber tracking using a modified FACT algorithm was performed until $100,000$ streamlines were reconstructed for each individual.

\subsection*{Structural Network Construction}

Anatomical scans were segmented using FreeSurfer \cite{Dale1999} and parcellated according to the Lausanne 2008 atlas included in the connectome mapping toolkit \cite{Hagmann2008}. A parcellation scheme including $234$ regions was registered to the B0 volume from each subject's DSI data. The B0 to MNI voxel mapping produced via QSDR was used to map region labels from native space to MNI coordinates. To extend region labels through the gray/white matter interface, the atlas was dilated by $4$mm. Dilation was accomplished by filling non-labeled voxels with the statistical mode of their neighbors' labels. In the event of a tie, one of the modes was arbitrarily selected. Each streamline was labeled according to its terminal region pair.

From these data, we built structural brain networks from each of the $61$ diffusion spectrum imaging scans.  Consistent with previous work 
\cite{Bassett2010,Bassett2011,Hermundstad2013,Hermundstad2014,Klimm2014,Gu2015,Muldoon2015,Muldoon2016,Sizemore2015}, we defined these structural brain networks from the streamlines linking $N=234$ large-scale cortical and subcortical regions extracted from the Lausanne atlas \cite{Hagmann2008}. We summarize these estimates in a weighted adjacency matrix $\mathbf{A}$ whose entries $A_{ij}$ reflect the structural connectivity between region $i$ and region $j$ (Fig.~\ref{fig1}A). 

Following \cite{Gu2015}, here we use an edge weight definition based on the \emph{quantitative anisotropy} (QA). QA is described by Yeh et. al (2010) as a measurement of the signal strength for a specific fiber population $\hat{a}$ in an ODF $\Psi(\hat{a})$ \cite{Yeh2010,Tuch2004}. QA is given by the difference between $\Psi(\hat{a})$ and the isotropic component of the spin density function (SDF, $\psi$) $\mbox{ISO}\left(\psi\right)$ scaled by the SDF's scaling constant. Along-streamline QA was calculated based on the angles actually used when tracking each streamline.  Although along-streamline QA is more specific to the anatomical structure being tracked, QA is more sensitive to MRI artifacts such as B1 inhomogeneity. QA is calculated for each streamline. We then averaged values over all streamlines connecting a pair of regions, and used this value to weight the edge between the regions. 

\subsection*{Resting state fMRI data}

As an interesting comparison to the computational model, we used resting state fMRI data collected from an independent cohort composed of 25 healthy right-handed adult subjects (15 female), with a mean age of 19.6 years (STD 2.06 year). All subjects gave informed consent in writing, in accordance with the Institutional Review Board of the University of California, Santa Barbara. Resting-state fMRI scans were collected on a 3.0-T Siemens Tim Trio scanner equipped with high performance gradients at the University of California, Santa Barbara Brain Imaging Center. A T2*-weighted echo-planar imaging (EPI) sequence was used (TR=2000 ms; TE=30 ms; flip angle=90°; acquisition matrix=64$\times$64; FOV=192 mm; acquisition voxel size = 3$\times$3$\times$3.5 mm; 37 interleaved slices; acquisition length=410s). 

We preprocessed the resting state fMRI data using an in-house script adapted from the workflows described in detail elsewhere \cite{baird2013medial,satterthwaite2013improved}.  The first four volumes of each sequence were dropped to control for instability effects of the scanner.  Slice timing and motion correction were performed in AFNIusing 3dvolreg and FreeSurfer’s BBRegister was used to co-register 
functional and anatomical spaces.  Brain, CSF, and WM masks were extracted, the time series were masked with the brain mask, and grand-mean scaling was applied.  The temporal derivative of the original 6 displacement and rotation motion parameters was obtained and the quadratic term was calculated for each of these 12 motion parameters, resulting in a total of 24 motion parameters which were regressed from the signal.  The principal components of physiological noise were estimated using CompCor (aCompCor and tCompCor) and these components were additionally regressed from the signal.  The global signal was not regressed.  Finally, signals were low passed below 0.1 Hz and high passed above 0.01 Hz in AFNI. To extract regional brain signals from the voxel-level time series, a mask for each brain region in the Lausanne2008 atlas was obtained and FreeSurfer was used to individually map regions 
to the subject space. A winner-takes-all algorithm was used to combine mapped regions into a single mask. The resulting signal for each region was then extracted in FreeSurfer using mrisegstats.

Following data preprocessing and time series extraction, we next turned to extracting observed brain states. Importantly, physiological changes relating to neural computations take place on a time scale much smaller than the time scale of BOLD image acquisition.  Thus, we treat each TR as representing a distinct brain state. To maximize consistency between the model-based and data-based approaches, we transformed the continuous BOLD magnitude values into a binary state vector by thresholding regional BOLD signals at 0. From the set of binary state vectors across all TRs, we defined activation rates in a manner identical to that described for the maximum entropy model data.

\subsection*{Defining an Energy Landscape}

We begin by defining a brain state both intuitively and in mathematical terms. A brain \emph{state} is a macroscopic 
pattern of 
BOLD activity across $K$ regions of the brain (Fig.~\ref{fig1}A). For simplicity, here we study the case in which each 
brain 
region 
$i$ can be either active ($\sigma_i=1$) or inactive ($\sigma_i=0$). Then, the binary vector $\bsigma = (\sigma_1, 
\sigma_2,\dots,\sigma_K)$ respresents a brain state configuration.

Next, we wish to define the energy of a brain state. We build on prior work demonstrating the neurophysiological relevance of maximum entropy models in estimating the energy of brain states in rest and task conditions \cite{Watanabe2013,Watanabe2014}. For a given state  $\bsigma$, we write its energy in the second order expansion:

\begin{equation} \label{eq:1}
E(\bsigma) = -\frac{1}{2}\sum_{i\neq j}J_{ij}\sigma_i \sigma_j - \sum_{i}J_{i} \sigma_{i},
\end{equation}
where $\mathbf{J}$ represents an interaction matrix whose elements $J_{ij}$ indicate the strength of the interaction between 
region 
$i$ and region $j$. If $J_{ij} > 0$, this edge $(i,j)$ decreases the energy of state $\bsigma$, while if $J_{ij} < 0$, 
this edge 
$(i,j)$ increases 
the 
energy of state $\bsigma$. The column sum of the structural brain network, $\J_{i} = \sum_{j}|J_{ij}|/\sqrt{K}$, is the strength of region 
$i$. In the thermodynamic equilibrium of the system associated with the energy defined in Eqn [\ref{eq:1}], the entropy of the system 
is maximized and the probability of the configuration $\bsigma$ is $P(\bsigma)\propto e^{-E(\bsigma)}$.

The choice of the interaction matrix is an important one, particularly as it tunes the relative contribution of edges to the system 
energy. In this study, we seek to study structural interactions in light of an appropriate null model. We therefore define the interaction 
matrix to be equal to the modularity matrix \cite{Newman2006} of the structural brain network: 
\begin{equation}\label{eq:j_ij}
\J_{ij} = \frac{1}{2m}(\A-\mathbf{p}\mathbf{p^T}/2m)_{ij}
\end{equation}  for $i\neq j$, where $\mathbf{A}$ is the adjacency matrix, $p_i = \sum_{i=1}^K A_{ij}$, and $2m = \sum_{j=1}^K 
p_j$.  
This choice 
ensures that any element $J_{ij}$ measures the difference between the strength of the edge $A_{ij}$ in the brain and the expected 
strength of that edge in an appropriate null model (here given as the Newman-Girvan null model \cite{clauset2004finding}). If the 
edge is 
stronger than expected, it will decrease the energy of the whole system when activated, while if the edge is weaker than expected, it 
will increase the energy of the whole system when activated. 

\subsection*{Discovering Local Minima}

The model described above provides an explicit correspondence between a brain's state and the energy of that 
state, in essence 
formalizing a multidimensional landscape on which brain dynamics may occur. We now turn to identifying and 
characterizing the local 
minima of that energy landscape (Fig.~\ref{fig1}C).  We begin by defining a local minimum: a binary state 
$\bsigma^* = 
(\sigma^*_1,\dots,\sigma^*_K)$ is a local minimum if $E(\bsigma) \geq E(\bsigma^*)$ for all vectors $\bsigma$ satisfying $||\bsigma-\bsigma^*||_1 = 1$, which means that the state $\bsigma^*$ realizes the lowest energy among its neighboring states within the closed unit sphere. We wish to collect all local minima in a matrix $\mathbf{\Sigma}^*$ with

\begin{equation}
\mathbf{\Sigma^*} = \begin{pmatrix}
  \sigma^*_{1,1} & \sigma^*_{1,2} & \cdots & \sigma^*_{1,N} \\
  \sigma^*_{2,1} & \sigma^*_{2,2} & \cdots & \sigma^*_{2,N} \\
  \vdots  & \vdots  & \ddots & \vdots  \\
  \sigma^*_{K,1} & \sigma^*_{K,2} & \cdots & \sigma^*_{K,N}
 \end{pmatrix}_{K\times N}
\end{equation}

\noindent where $N$ is the number of local minima and $K$ is the number of nodes in the structural brain 
network (or 
equivalently 
the cardinality of the adjacency matrix $\mathbf{A}$).

Now that we have defined a local minimum of the energy landscape, we wish to discover these local minima 
 given the 
pattern of white matter connections represented in structural brain networks. To discover local minima of $E(\bsigma)$, we first note 
that the total number of states $\bsigma = (\sigma_1, \dots,\sigma_K)$ is $2^K$, which -- when $K=234$ -- 
prohibits an exhaustive analysis of all possibilities. Moreover, the problem of finding the ground state is an NP-complete 
problem 
\cite{cipra2000ising}, and thus it is 
unrealistic to expect to identify all local minima of a structural brain network. We therefore choose to employ a clever heuristic to 
identify local minima. Specifically, we combine the Metropolis Sampling method \cite{metropolis1953equation} and a steep search 
algorithm using gradient descent methods. We identify a starting position by choosing a state uniformly at random from the set of all 
possible states. Then, we step through the energy landscape via a random walk driven by the Metropolis Sampling criteria (see 
Algorithm~\ref{al:1}). At each point on the walk, we use the steep search algorithm to identify the closest local minimum.
%
%

\begin{algorithm}[H] \label{al:1}
	Let $\bsigma^j$ be the vector obtained by changing the value of the $j$-th entry of $\bsigma$\;
	\For {t = 1 : N}{
		Randomly select an index $j\in\{1,...,K\}$\;
		Set $\tilde{\bsigma}_t = \bsigma_t=\bsigma_{t-1}^j$ with probability $p = \min(1, e^{-\beta*(E(\tilde{\bsigma}) 
		-E( \bsigma_{t-1}))})$ and  $\tilde{\bsigma}_t  = \bsigma_t =\bsigma_{t-1}$ otherwise\;
		\While{${\tilde{\bsigma}_t}$ is not a local minimum}{
			Set $j^*= \arg\min_j  E(\tilde{\bsigma}_t^{j})$\;
			Set $\tilde{\bsigma_t} = \tilde{\bsigma}_t^{j^*}$\;
		}
		Set $\bsigma^*_t = \tilde{\bsigma_t}$\;
	}
	\caption{Heuristic Algorithm to Sample the Energy Landscape to Identify Local Minima.}
\end{algorithm}

Here, $\bsigma_{1},\bsigma_{2},\dots, \bsigma_{N}$ are the sampled states, ${\bsigma^*_1}, {\bsigma^*_2},\dots, 
{\bsigma^*_N}$ are the sampled local minima, and $\beta$ is the temperature parameter which can be absorbed in 
$E(\bsigma)$. In 
the context of any sampling procedure, it is important to determine the number of samples necessary to 
adequately cover the space. Theoretically, we wish to identify a number of samples $N$ following which the 
distribution of energies 
of the local minima remains stable. Practically speaking, we choose 4 million samples in this study, and demonstrate the stability of 
the energy distribution in the Supplement. A second important consideration is to determine the initial state, that is the state from 
which the random walk begins. Here we choose this state uniformly at random from the set of all possible states. However, this 
dependence on a uniform probability distribution may not be consistent with the actual probability of states in the energy landscape. 
We therefore must ensure that our results are not dependent on our choice of initial state. To ensure independence from the initial 
state, we dismiss the first $30,000$ local minima identified, and we demonstrate in the Supplement that this procedure ensures our 
results are not dependent on the choice of initial state. 

\subsection*{Characterizing Local Minima}

Following collection of local minima, we wished to characterize their nature as well as their relationships to one 
another. First, we 
estimated the radius of each local minimum as the Hamming distance from the minimum to the closest sampled 
point on energy 
landscape.  Next, calculated the Hamming distance from each local minimum to the first sampled local 
minimum, a second 
quantification of the diversity of the energy lanscape that we traverse in our sampling. Finally, we quantify how 
diverse the observed 
local minima are by calculating the pairwise normalized mutual information\cite{manning2008introduction} of each 
pair of local minima. 

Next, we wished to understand the role of different regions and cognitive systems in the minimal energy states. Cognitive systems are sets of brain regions that show coordinated activity profiles in resting state or task-based BOLD fMRI data \cite{Mattar2015}. They include the visual, somatosensory, motor, auditory, default mode, salience, fronto-parietal, cingulo-opercular, dorsal and ventral attention systems, as well as subcortical areas. Here, the specific association of regions of interest to cognitive systems are exactly as listed in \cite{Gu2015} and based originally on results described in \cite{Power2011}. We characterize the roles of these systems in the local minima by assessing their \emph{activation rates}, as well as the \emph{utilization energies} required for communication within and between systems.

\subsection*{Activation Rates} Intuitively, we define the activation rate of a node $i$ as the average activation state of that node over all the local minima. Formally, the activation rate for region $i$ is defined as
\begin{equation}
r_i = \frac{\sum_{l=1}^N \sigma^*_{il}}{N},
\label{eq:3}
\end{equation}
where $l$ indexes over states, and recall $N$ is the number of local minima. The computed activation rate offers a prediction of which regions are more \emph{versus} less active across the local minima (that is, the brain's locally ``stable " states), and can be directly compared with the resting state activation rate estimated from empirical fMRI data.

\subsection*{Utilization Energies} To complement the information provided by the activation rates, we also defined the 
energetic costs associated with utilizing within- and between-systems interactions. We note that each cognitive system is a 
subnetwork of the whole brain network. We use the index set $\mathcal{I}$ to indicate the set of nodes associated with the cognitive 
system, and thus $|\mathcal{I}|$ gives the number of nodes in the system. Then, for a given state $\bsigma$, the within-system energy 
measures the cost associated with the set of interactions constituting the subnetwork. The between-system energy measures the cost 
associated with the set of interactions between the subnetwork and all other nodes in the whole network. Formally, we define

\begin{eqnarray*}
E^W(\bsigma) &=& -\frac{1}{2|\mathcal{I}|(|\mathcal{I}|-1)}\left(\sum_{i\neq j,i,j \in\mathcal{I}}J_{ij}\sigma_i \sigma_j \right)\\
E^{B}(\bsigma) &=& -\frac{1}{2|\mathcal{I}|(K-|\mathcal{I}|)}\left(\sum_{i\in\mathcal{I}, j\notin\mathcal{I}} J_{ij} \sigma_i \sigma_j \right)
\end{eqnarray*}

\noindent where $E^{W}$ measures the within-system energy, $E^{B}$ measures the between-system energy, and the normalization coefficients $1/(|\mathcal{I}||\mathcal{I}|-1|)$, $1/(|\mathcal{I}|(K-|\mathcal{I}|))$ are chosen by considering the number of the corresponding interactions.

\subsection*{Permutation Tests for State Association}
For a given local minimum configuration $\bsigma^*$, we associate it with system $i_{\sigma^*}$,  
\begin{equation*} 
i_{\sigma^*} = \arg\max_i \mathrm{NMI}(\bsigma^*, \bsigma^{\mathrm{sys}}_i),
\end{equation*}
where $\bsigma^{\mathrm{sys}}_i$ is the configuration pattern of system $i$ such that the corresponding regions for system $i$ are activated and the others not and where ``NMI'' refers to the Normalized Mutual 
Information\cite{manning2008introduction}, which is used to measure the similarities between the two states. 
To obtain the null distribution, for each local 
minimum configuration $\bsigma^*$ in the 
collection $\mathbf{\Sigma^*}$,  we permute the configuration at each position of $\bsigma^*$  to achieve a 
random configuration 
with the same activation rate, 
and we then compute the associated percentage in each system. Then we repeat this procedure to generate $N$ samples and 
construct the null distribution of the probability of being configured as each system pattern. Considering the large size of the state 
collection, the variance of the samples in the null distribution will be small. We pick $N=50$ here. See Fig.~\ref{fig4} D for the 
results.

\section*{Results}

\subsection*{Local Minima in the Brain's Energy Landscape}

By sampling the energy landscape of each structural connectivity matrix, we identified an average of approximately 450 local minima or low energy brain states: binary vectors indicating the pattern of active and inactive brain regions (see Methods). On average across the 61 scans, 144 brain regions were active in a given local minimum, representing 61.70\% of the total (standard deviation: 6.04\%).  This large percentage suggests that the brain may utilize a broad set of rich and diverse activations to perform cognitive functions \cite{Chai2016}, rather than activation patterns localized to small geographic areas.

\begin{figure}[!h]
\includegraphics[width=4in]{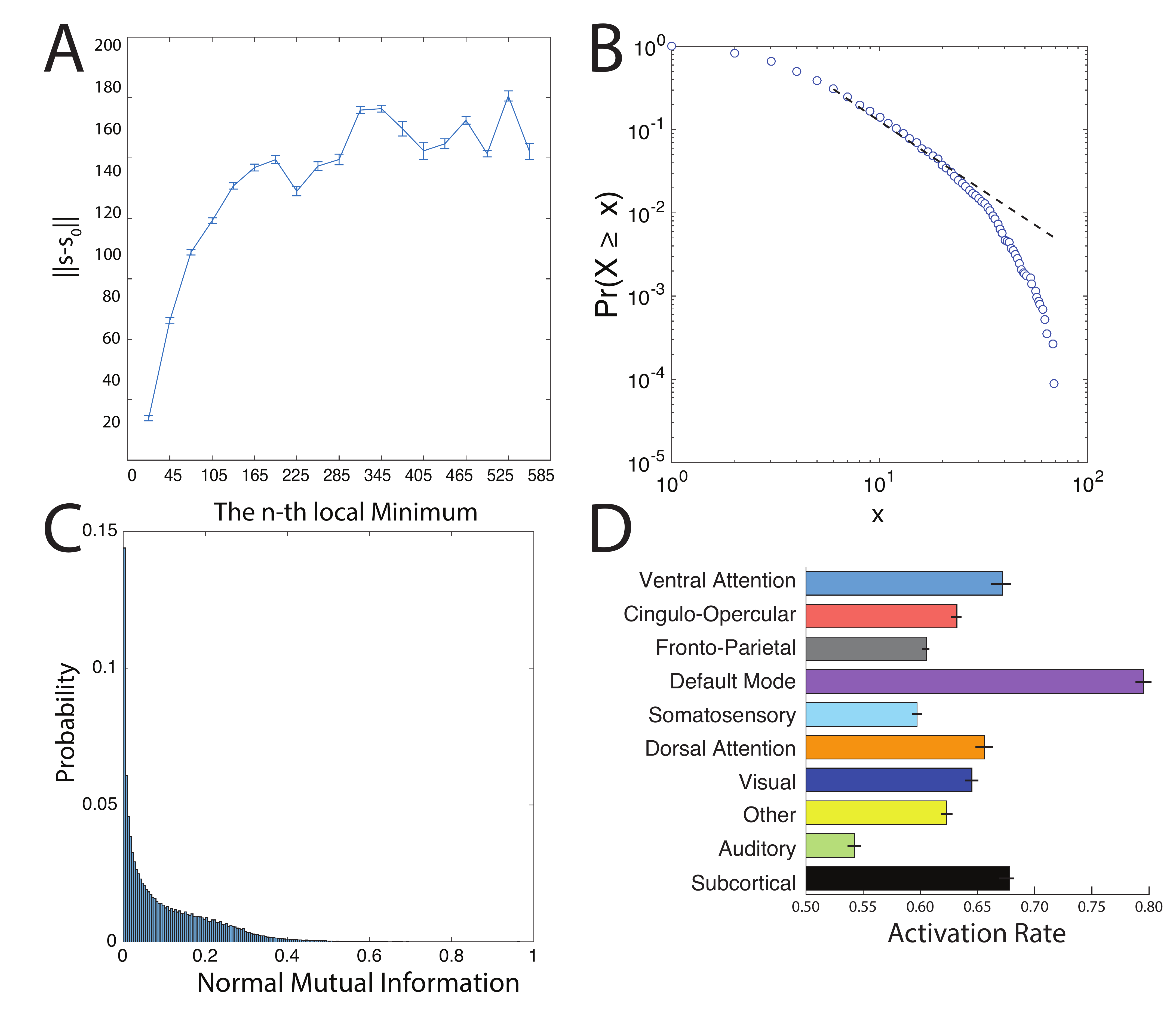} 
\caption{\textbf{Simulated Activation Rates.}\emph{(A)} The distribution of distances from the first local minimum to other local minima. Each point and error-bar is calculated across a bin of 30 minima; error bars indicate standard error of the mean over the 30 minima. \emph{(B)}  The probability distribution of the radius of each local minimum is well-fit by a power-law. The radius of a local minimum is defined as its distance to the closest sampled point on the energy landscape.  \emph{(C)} The distribution of the pairwise 
normalized mutual information between all pairs of local minima.  \emph{(D)} Average activation rates for all 14 \emph{a priori} defined cognitive systems \cite{Power2011}. Error bars indicate standard error of the mean across subjects. }
  \label{fig2}
 \end{figure}

To quantify this diversity, we examined the location of minima on the energy landscape, the size of the basins surrounding the minima, and the mutual information between minima. First, we estimated the distance from the first local minima identified to all subsequent minima (see  Methods; Fig.~\ref{fig2}A). We observe an order of magnitude change in the distance between the first and second local minima, and the first and last local minima, suggesting that local minima span a broad geographic domain in the energy landscape. Interestingly, these minima differ not only in their location on the energy landscape, but also in the size of the basins surrounding them. We estimate basin size by calculating the radius of each local minimum (see Methods) and show that the distribution of radii follows a power-law, with the majority of minima displaying a small radius, and only a few minima displaying a large radius (Fig.~\ref{fig2}B). More specifically, we fit the function $P(r)=Cr^{-\alpha}$ -- where $C$ is a constant -- to the data using a statistically principled approach \cite{Clauset2009,Virkar2014}. We identified an $\alpha = 2.6300$ for $r > 6$, and the $p$-value for the goodness of fit was $p < 1 \times 10^{-5}$ indicating that the power law was an accurate fit to the data.  As a final quantification of minima diversity, we estimated the normalized mutual information between every pair of local minima, as an intuitive measurement of the similarity between anatomical compositions of the minima. We observe that the probability distribution of normalized mutual information between minima pairs is heavy tailed, indicating that most minima pairs display very dissimilar anatomical compositions, and only a few minima pairs display similar anatomical compositions (Fig.~\ref{fig2}C).

From a neurophysiological perspective, it is also important to note that these local minima displayed significant local structure. Specifically, we found that regions within known cognitive systems tended to be active together. The probability that regions were co-active is 48.22\%, which was significantly greater than that expected in a null distribution (associated $p-value$ was $p <1 \times 10^{-5}$; see 
Methods). These results indicate that the structural connectivity between brain regions, and the assumption of energy conservation, predict that regions that belong to the same cognitive system will tend to be co active with one another during diverse cognitive functions. Indeed, these predictions are consistent with previous studies of functional neuroimaging data demonstrating that groups co-active regions tended to align well with known cognitive systems \cite{Crossley2013,Crossley2014}.

\subsection*{Activation Rates of Cognitive Systems}
Given the alignment of activation patterns with cognitive systems, we next asked whether certain cognitive systems were activated more frequently than others. To address this question, we studied the \emph{activation rate} of each cognitive system, which measures how frequently the regions in the cognitive system participated in the set of states identified as local minima. Intuitively, if the activation rate is high, the system is more likely to be active in diverse brain states. We observed that systems indeed showed signficantly different activation rates (Fig.~\ref{fig2}D). Sensorimotor systems (auditory, visual, somatosensory) tended to display the lowest activation rates, followed by higher order cognitive systems (salience, attention, fronto-parietal, and cingulo-opercular), and subcortical structures. The system with the largest activate rate was the default mode system, suggesting that activation of this system is particularly explicable from structural connectivity and the assumption of energy conservation. The unique role of the default mode system is consistent with predictions from network control theory that highlight the optimal placement of default mode regions within the network to maximize potential to move the brain into many easily reachable states with minimal energetic costs \cite{Gu2015}.

It is important to determine whether this activation rate is driven by simple properties of the structural connectivity matrix that do not 
depend on assumptions of energy conservation. To address this question, we next assessed the relationship between a simple summary 
statistic of the structural connectivity matrix -- the \emph{strength}, or weighted degree, of a brain region -- and the predicted activation rate 
drawn from the maximum entropy model. We observed that the activation rate was not well predicted by the weighted degree on 
average over brain regions (see Supplement). These data suggest that the additional assumption of energy conservation produces a set of 
brain states that cannot be predicted from simple statistics of structural connectivity alone.

\subsection*{Relating Predicted Activation Rates to Rates Observed in Functional Neuroimaging Data}

Before exercising the model further to explore how energy is utilized in the brain, we wished to quantify the relationships between the theoretically predicted activation rates, and activation rates observed empirically in functional neuroimaging data. Specifically, we studied fMRI data acquired in a separate set of healthy adult human subjects. 

\begin{figure}[!h]
\includegraphics[width=145mm]{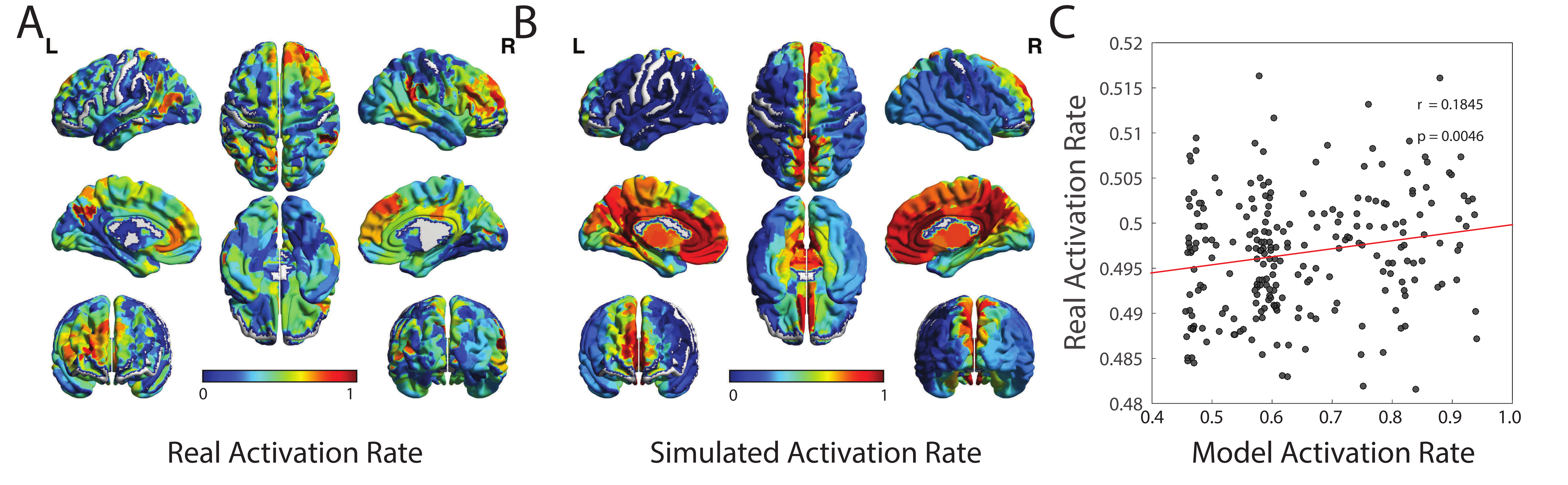} 
\caption{\textbf{Validating Predicted Activation Rates in Functional Neuroimaging Data.} \emph{(A)} From resting state BOLD data acquired in an independent cohort, we estimated the true activation rate by transforming the continuous BOLD magnitudes to binary state vectors by thresholding the signals at 0 (see Methods). We use these binary state vectors to estimate the activation rates of each brain region across the full resting state scan. Here we show the mean activation rate of each brain region, averaged over subjects.  \emph{(B)} For comparison, we also show the mean predicted activation rate estimated from the local minima of the maximum entropy model, as defined in Equation~[\ref{eq:3}], and averaged over subjects. \emph{(C)} We observe that the activation rates estimated from resting state fMRI data are significantly positively correlated with the activation rates estimated from the local minima of the maximum entropy model (Pearson's correlation coefficient $r=0.18$, $p=0.0046$). Each data point represents a brain region, with either observed or predicted activation rates averaged over subjects.}
\label{fig3}
\end{figure}

Next, we wished to directly quantify similarities between the predicted activation rates and those observed empirically in resting state 
fMRI. In the resting state data, we observed that highly active regions were located in broad swaths of frontal and parietal cortex, as 
well as medial prefrontal, precuneus, and cingulate (Fig.~\ref{fig3}A). This pattern of high activation is consistent with the so-called 
``default-mode'' of resting state brain function \cite{raichle2001default}. In our maximum entropy model, we observed that the areas 
predicted to have high activation rates show a broad similarity to those observed empirically in the resting state (Fig.~\ref{fig3}B). 
Indeed, we observed that the empirical resting activation rate of brain regions is significantly correlated with the activation rate 
predicted from the maximum entropy model (Fig.~\ref{fig3}C; Pearson correlation coefficient $r=0.18$, $p=0.0046$). These results 
suggest that the modeling framework we use here has significant similarities to observable features of resting state brain dynamics. 
However, it is important to mention that there are also noticeable differences between the two maps: the predicted activation rates 
are strong along the medial wall, while the resting state activation rates extend to larger sections of lateral cortices.

\subsection*{Utilization Energies of Cognitive Systems}

\begin{figure}[!h]
\includegraphics[width=4.5in]{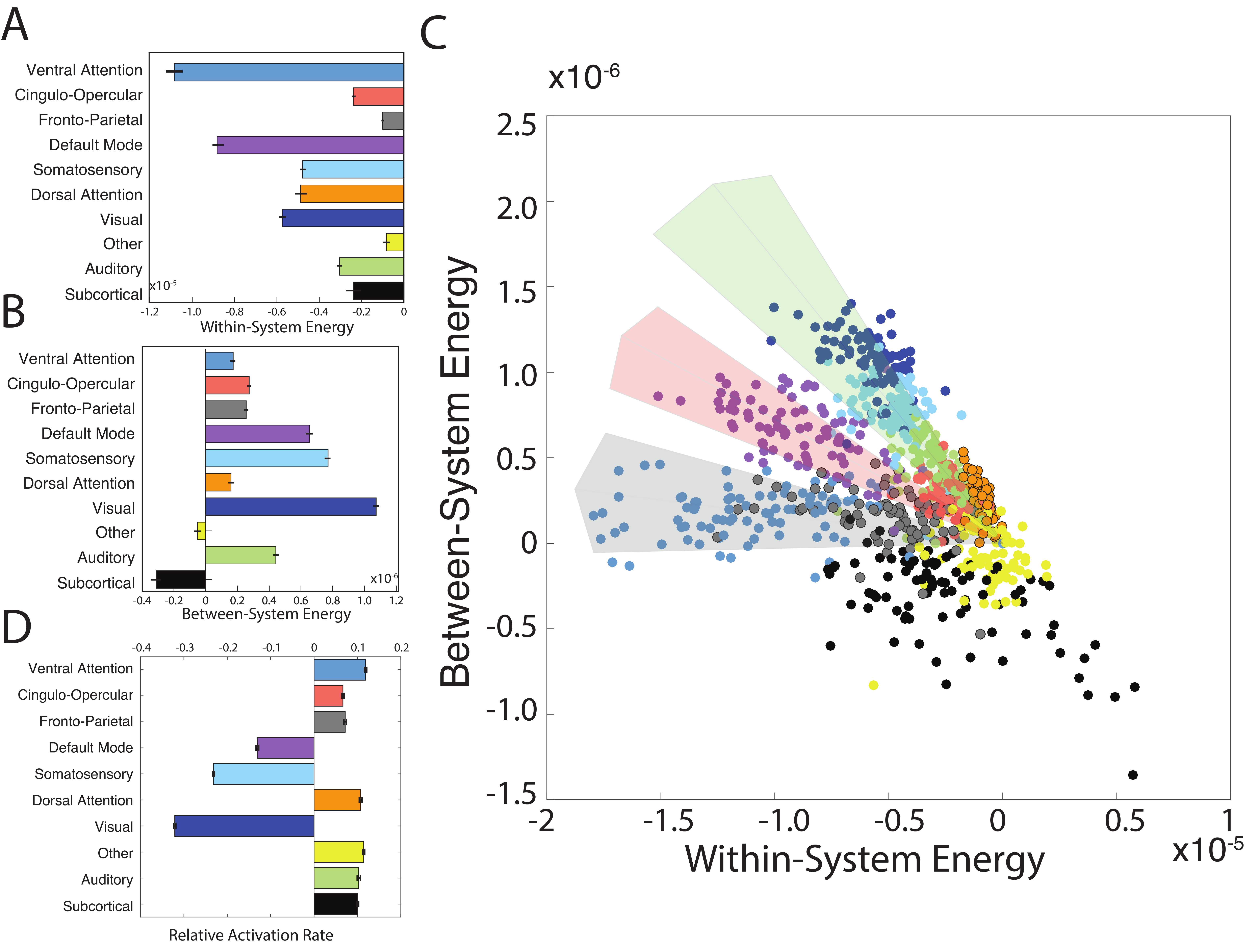}
\caption{\textbf{Utilization Energies of Cognitive Systems.} \emph{(A)} Average within-system energy of each cognitive system; error bars indicate standard error of the mean across subjects. \emph{(B)} Average between-system energy of each cognitive system; error bars indicate standard error of the mean across subjects.  \emph{(C)} The 2-dimensional plane mapped out by the within- and between-system energies of different brain systems. Each data point represents a different brain region, and visual clusters of regions are highlighted with lightly colored sectors. The sector direction is determined by minimizing the squared loss in point density of the local cloud and the width is determined by the orthogonal standard derivation at the center along the sector direction. In this panel, all data points represent values averaged across 
subjects. \emph{(D)} The percentages of minima displaying preferential activation of each system; each minima was assigned to the system which whom it shared the largest normalized mutual information. Errorbars indicate the differences between the observed percentages and those of the null distribution with random activation patterns across regions.}
  \label{fig4}
 \end{figure}

We next turned to exercising our model to further understand the potential constraints on brain state dynamics. Specifically, we asked how cognitive systems utilized the minimal energy presumably available to them.  Intuitively, this question encompasses both how energy is utilized by \emph{within}-system interactions, and how energy is utilized by \emph{between}-system interactions. We therefore defined the within-system energy, which measures the cost associated with the set of interactions constituting the cognitive system, and the between-system energy, which measures the cost associated with the set of interactions between cognitive systems. We observed a fairly strong dissociation between these two variables: cognitive systems that display a large within-system energy are not necessarily those that display a large between-system energy (see Fig.~\ref{fig4}A and B). Indeed, within- and between-system energies are not significantly correlated across systems (Pearson's correlation coefficient $r=0.2287$ and $p=0.5250$), suggesting that these two variables offer markers of distinct constraints.

Moreover, we observed that the 2-dimensional plane mapped out by the within- and between-system energies of all brain regions revealed the presence of 4 surprisingly distinct clusters (Fig.~\ref{fig4}C) that are not explicable by simple statistics such as network degree (see Supplement). Each cluster represents a unique strategy in energy utilization that is directly reflected in its activation pattern; in other words, each cluster offers a distinct balance between the energetic costs of within-system interactions and the energetic costs of between-system interactions. The central cluster, displaying high within-system energies but low between-system energies, is composed of subcortical and fronto-parietal systems. A high between-system energy cone emanating from this central cluster is composed of predominantly primary and secondary sensorimotor cortices in somatosensory, visual, and auditory systems. A second cone emanating from the central cluster with a slightly lower between-system energy is composed predominantly of regions in the default mode system. The final cone emanating from the central cluster with between-system energies near zero is composed predominantly of regions in the dorsal and ventral attention systems. These results suggest that sensorimotor, default mode, attention, and cognitive control circuits display differential preferences for energy utilization: regions in attentional systems share less energy with other networks than regions in sensorimotor systems, while the default mode maintains an intermediate balance.

The clear differences in the energies utilized by different cognitive systems and by between-system interactions begs the question of 
whether the brain cares about these energies. Does the brain prefer smaller within-system energies, smaller between-system 
energies, or some balance between the two? To address this question, we studied the ensemble of local minima, and asked which 
systems were commonly expressed. Specifically, for each local minimum, we determined which system was most activated, assigned 
the minimum to that system, and performed this assignment for all minima. We observed that 3 systems were represented at higher 
percentages than expected in a permutation-based null model (see Methods): the default mode system, the 
visual system, and the somatosensory system (Fig.~\ref{fig4}D). Importantly, these three systems represent the systems with the 
highest 
between system energies (Fig.~\ref{fig4}C), but are indistinguishable from other cognitive systems in terms of within-system energy. 
These results suggest that the brain may prefer high integration between systems over low integration, and that the constraint of 
between-system energies is more fundamental to brain function than the constraint of within-system energies.

\section*{Discussion}

In this paper, we address the question of how large-scale brain circuitry and distinct energetic constraints produce whole-brain patterns of activity. We build our approach on a maximum entropy model of brain dynamics that is explicitly informed by estimates of white matter microstructure derived from deterministic tractography algorithms. The model allows us to study minimal energy states, which we observe to be composed of co-activity in local spatially-contiguous sets of brain regions reminiscent of cognitive systems. These systems are differentially active, and activity patterns are significantly correlated with the observed activation rate measured in a separate resting state fMRI data set. Finally, we exercise this model to ask how cognitive systems utilize the minimal energy presumably available to them. We find that the energy utilized within and between cognitive systems distinguishes 4 classes of energy utilization dynamics, corresponding to sensorimotor, default mode, attention, and cognitive control functions. These results suggest that diverse cognitive systems are optimized for differential contributions to integrated \emph{versus} segregated function via distinct patterns of energy utilization. More generally, the results highlight the importance of considering energetic constraints in linking structural connectivity to observed  dynamics of neural activity.

\subsection*{The Role of Activation vs. Connectivity in Understanding Brain Dynamics}

As the interest in understanding structural brain connectomics has blossomed in the last several years \cite{Sporns2005,Sporns2011}, it has not been accompanied by an equally vivid interest in linking subsequent insights to the more traditional notions of brain activation profiles \cite{Bassett2015}. Indeed, the fields of systems and cognitive neuroscience have instead experienced a pervasive divide between the relatively newer notions of \emph{connectome mapping} and the relatively traditional yet highly effective notions of brain mapping \cite{Sporns2015}, which have led to powerful insights into neural function in the last quarter century \cite{Zeki2005}. This divide is at least in part due to the fact that graph theory and network-based methods on which the field of connectomics is based have few tools available to link node properties (activity) with edge properties (connectivity) \cite{Newman2015}. While technically explicable, however, the conceptual divide between these fields can only lead to their detriment, and synergistic efforts are necessary to develop a language in which both activity and connectivity can be examined in concert \cite{Bassett2015}. This study offers one explicit mathematical modeling framework in which to study the relationships between activation profiles across the whole brain and underlying structural connectivity linking brain regions. Complementary approaches include the model-based techniques formalized in the publicly available resource \emph{The Virtual Brain} \cite{Shirner2015,Ritter2013}.

\subsection*{Co-activation Architecture}
In this study, we observed that brain regions affiliated with known cognitive systems -- including somatosensory, visual, auditory, default mode, dorsal and ventral attention, fronto-parietal, and cingulo-opercular -- also tend to be active together with one another in low energy brain states. Indeed, these theoretical results are consistent with previous studies of functional neuroimaging data demonstrating that groups of co-active regions tended to align well with known cognitive systems \cite{Crossley2013,Crossley2014}. This correspondence is particularly interesting when one considers how these cognitive systems were initially defined: and that is, based on strong and dense functional connectivity \cite{Power2011}. Thus, our results point to a fundamental mapping between activity and connectivity: regions that are active together tend to be functionally connected together.  The presence of such a map has been empirically observed in the resting state (though not in task \cite{Bassett2015}), in both healthy controls and in people with schizophrenia where the map appears to be fundamentally altered in its nature \cite{Bassett2012,Zalesky2012,Yu2013}. Here we offer a mechanism for this mapping based on a combined consideration of energy- and connectivity-based constraints.

\subsection*{Critical Importance of Energy Constraints}
The quest to understand and predict brain dynamics from the architecture of underlying structural connectivity is certainly not a new one. In fact, there have been concerted efforts over the last decade and more to identify structural predictors of the resting state BOLD signal. Seminal contributions have included the observations of statistically significant correlations between structural connectivity estimated from diffusion imaging data and functional connectivity estimated from fMRI \cite{Honey2009}, as well as extensions of these correlations that account for long distance paths along white matter tracts \cite{Goni2014} and spectral properties of structural matrices \cite{Becker2016}. The question of how brain structure constrains a wide range of brain states (beyond simply the resting state) is a very open area of inquiry. Moreover, this question is particularly challenging to address with empirical data because it is difficult to obtain data from humans in more than a handful of task states \cite{Cole2014}.  For this reason, computational models play a very important role in offering testbeds for the development of theories linking structure to ensembles of brain states, which can in turn offer testable predictions. Our results suggest that an understanding of the relationship between brain structure and function is perhaps ill-constrained when examining connectivity alone. The additional assumption of energy conservation produces a set of brain states that cannot be simply predicted from statistics of structural connectivity, perhaps offering a mechanism for the large amount of unexplained variance in prior predictions \cite{Honey2009,Goni2014}. 

\subsection*{Methodological Considerations}
Our results are built on the formalism of the maximum entropy model, which is predicated on pairwise statistics \cite{Bialek2012}. However, emerging evidence suggests that some neurophysiological phenomenon are better studied in the framework of simplicial complexes rather than dyads \cite{Giusti2016}. For example, integrate and fire neurons exposed to common fluctuating input display strong beyond-pairwise correlations that cannot be captured by maximum entropy models \cite{Leen2015}. Similar arguments can also be made for co-activation patterns in BOLD fMRI \cite{Crossley2013,Crossley2014,Fox2014}. It will be interesting in future to determine the role energetic and structural constraints on observed higher-order functional interactions during human cognitive function.

A second important consideration is that the maximum entropy model is appropriate for systems at equilibrium. Therefore, the local minima identified may not accurately represent the full class of states expected to be elicited by daily activity.  Instead, the local minima identified here are expected to more accurately represent the set of states expected to appear as a person rests in the so-called \emph{default mode} of brain function, which is thought to lie near a stable equilibrium \cite{Deco2012}. Such an interpretation is consistent with our findings that the activation rates predicted by the maximum entropy model are strongly correlated with the activation rates observed in resting state fMRI data.

\section*{Conclusion}

The analyses presented in this study produce information regarding an underlying energy landscape through which the brain is predicted to move. The existence of such a landscape motivates the very interesting question of how the brain transitions between states. In sampling this landscape, we have used a simple random walk in an effort to extract a large ensemble of possible brain states, measured by local minima. However, the question of which walks a healthy (or diseased) brain might take through this landscape remains open. Such walks or dynamical trajectories may be determined by energetic inputs to certain regions of the brain \cite{Gu2015}, either by external stimuli or by neuromodulation \cite{Muldoon2016}. In this context, network control theory may offer explicit predictions regarding the optimal dynamic trajectories that the brain may take through a set of states to move from an initial state to a target state with little energetic resources \cite{Gu2015,Pasqualetti2014}. In addition to inputs to single regions, changes in a cognitive task -- for example elicited by task-switching paradigms -- may also drive a specific trajectory of brain states. Indeed, it is intuitively plausible that the asymmetric switch costs observed between cognitively effortful and less effortful tasks \cite{Wu2015,Davidson2006} may be explained by characteristics of the energy landscape defined by structural connections between task-activated brain regions. 

\newpage
\section*{Supporting Information}

\paragraph*{S1 Fig.}
\label{S1_Fig}
{\bf Stability of the Energy Distribution with respect to the Number of Samples.} We plot the probability distribution of the energy for the first 2000, 4000, 6000 and 8000 samples. We observe that the shapes of the probability distributions are qualitatively consistent. We confirm this qualitative observation with Kolmogorov-Smirnov tests (see Supplemental text).

\paragraph*{S2 Fig.}
\label{S2_Fig}
{\bf Activation Rate is Poorly Predicted by Regional Degree and Energy.} \emph{(A)} Scatterplot of weighted regional degree \emph{versus} activation rate. \emph{(B)} Scatterplot of regional energy \emph{versus} activation rate. We observe that the activation rate is not well predicted by either regional energy or weighted degree. 

\paragraph*{S3 Fig.}
\label{S3_Fig}
{\bf Relationship Between Utilization Energies and Degree.}  \emph{(A)} Scatterplot of within-system connectivity \emph{versus} between-system connectivity, for individual brain regions that are color-coded by cognitive system. \emph{(B)} The same data presented in panel \emph{(A)} except only for the default mode, attention, and task control systems, demonstrating the indistinguishability of default mode and attention systems. \emph{(C)} Scatterplot of the between- and within- system energy. \emph{(D)} The same data presented in panel \emph{(C)} except only for the detault mode, attention, and task control systems. We observe that cognitive systems are more clearly separated in the 2-dimensional space of the within- and between-system energies than in the 2-dimensional space of the within- and between-system connectivity. Across all four panels, data points indicate brain regions, and color of data points indicates which cognitive system that region is affiliated with (see legend for map from color to cognitive system).

\paragraph*{S4 Fig.}
\label{S4_Fig}
{\bf Simulated Activation Rate is Significantly Correlated with the Rates Observed in Resting State Functional Neuroimaging Data.}  We observe that the activation rates estimated from resting state fMRI data are significantly positively correlated with the activation rates estimated from the local minima of the maximum entropy model (Pearson's correlation coefficient $r=0.17$, $p=0.0094$). Each data point represents a brain region, with either observed or predicted activation rates averaged over subjects.

\paragraph*{S1 Appendix.}
\label{S1_Appendix}
{\bf S1 Appendix.} Supplementary Information for ``The Energy Landscape of Neurophysiological Activity Implicit in Brain Network Structure''.

\section*{Acknowledgments}
We thank Ankit Khambhati, Lia Papadopoulus, and Evelyn Tang for helpful comments on an earlier fersion of this manuscript. D.S.B., S.G., and R.F.B. acknowledge support  from the John D. and Catherine T. MacArthur Foundation, the Alfred P. Sloan Foundation, the Army Research Laboratory and the Army Research Office through contract numbers W911NF-10-2-0022 and W911NF-14-1-0679,the National Institute of Mental Health (2-R01-DC-009209-11), the National Institute of Child Health and Human Development (1R01HD086888-01), the Office of Naval Research, and the National Science Foundation (BCS-1441502, BCS-1430087, and PHY-1554488). FP acknowledges support from BCS-1430280. B.B. was supported by the National Institutes of Health under Ruth L. Kirschstein National Research Service Award F32NS089348 from the NINDS. 

\nolinenumbers

%
%

\newpage
\bibliographystyle{plos2015}
\bibliography{energyLandscape}

\end{document}